\documentclass[times,twocolumn,final]{elsarticle}

\usepackage{medima}
\usepackage{framed,multirow}

\usepackage{amssymb}
\usepackage{latexsym}

\usepackage{url}
\usepackage{xcolor}

\usepackage{hyperref}

\usepackage{amsmath,amssymb,amsfonts}
\usepackage{graphicx}
\usepackage{textcomp}
\usepackage[ruled]{algorithm2e}
\usepackage{algpseudocode}
\usepackage{multirow}

\usepackage{subfig}
\usepackage{mathtools}

\definecolor{newcolor}{rgb}{.8,.349,.1}

\journal{Medical Image Analysis}

\begin{document}

\verso{Shaheer U. Saeed \textit{et~al.}}

\begin{frontmatter}

\title{Image quality assessment for machine learning tasks using meta-reinforcement learning}%

\author[1]{Shaheer U. \snm{Saeed}\corref{cor1}}
\cortext[cor1]{Corresponding author: shaheer.saeed.17@ucl.ac.uk}
\author[1,2]{Yunguan \snm{Fu}}
\author[3,4]{Vasilis \snm{Stavrinides}}
\author[1]{Zachary M. C. \snm{Baum}}
\author[1]{Qianye \snm{Yang}}
\author[5]{Mirabela \snm{Rusu}}
\author[6]{Richard E. \snm{Fan}}
\author[5,6]{Geoffrey A. \snm{Sonn}}
\author[7]{J. Alison \snm{Noble}}
\author[1]{Dean C. \snm{Barratt}}
\author[1,7]{Yipeng \snm{Hu}}

\address[1]{Centre for Medical Image Computing, Wellcome/EPSRC Centre for Interventional \& Surgical Sciences, and Department of Medical Physics \& Biomedical Engineering, University College London, London, UK} 
\address[2]{InstaDeep, London, UK}
\address[3]{Division of Surgery \& Interventional Science, University College London, London, UK}
\address[4]{Department of Urology, University College Hospital NHS Foundation Trust, London, UK}
\address[5]{Department of Radiology, Stanford University, Stanford, California, USA}
\address[6]{Department of Urology, Stanford University, Stanford, California, USA}
\address[7]{Department of Engineering Science, University of Oxford, Oxford, UK}

\received{1 May 2013}
\finalform{10 May 2013}
\accepted{13 May 2013}
\availableonline{15 May 2013}
\communicated{S. Sarkar}

\begin{abstract}
In this paper, we consider image quality assessment (IQA) as a measure of how images are amenable with respect to a given downstream task, or \textit{task amenability}. When the task is performed using machine learning algorithms, such as a neural-network-based \textit{task predictor} for image classification or segmentation, the performance of the task predictor provides an objective estimate of task amenability. In this work, we use an \textit{IQA controller} to predict the task amenability which, itself being parameterised by neural networks, can be trained simultaneously with the task predictor. We further develop a meta-reinforcement learning framework to improve the adaptability for both IQA controllers and task predictors, such that they can be fine-tuned efficiently on new datasets or meta-tasks. We demonstrate the efficacy of the proposed task-specific, adaptable IQA approach, using two clinical applications for ultrasound-guided prostate intervention and pneumonia detection on X-ray images.
\end{abstract}


\end{frontmatter}


\section{Introduction}
\label{sec:introduction}

\subsection{Image quality assessment}

Medical imaging is used extensively for diagnostic and therapeutic procedures in medicine, whether they be interventional or non-interventional in nature. Several such diagnostic, navigational or therapeutic tasks in the clinical workflow rely on medical images where they inform the clinician's judgement, directly or via derived measurements. Medical imaging is increasingly being used as a navigational aid to guide surgical and other interventional procedures, such as for prostate biopsies \citep{brown_prostate_biopsy}, liver resections \citep{simpson_liver_resection}, and brain resections \citep{kondziolka_brain_resection}. Treatment planning, for example radiotherapy planning, relies heavily on pre-operative medical images \citep{dirix_mr_radiotherapy, liney_mr_radiotherapy_q}. Moreover, imaging is commonly used for diagnostic clinical tasks whether the task is performed manually by humans or automated using computer aided diagnosis. The use of chest radio-graphs for diagnosis of lung diseases \citep{doi_cad}, computed tomography (CT) or magnetic resonance (MR) scans for diagnosis of brain diseases \citep{doi_cad} and ultrasound (US) for diagnosis of uterine diseases \citep{dueholm_tvus}, are all common examples where diagnosis relies heavily or solely on medical images.

The performance of clinical tasks that rely on medical imaging can be adversely impacted by the image quality of the images being used \citep{qa_review}. Image quality assessment (IQA) is an effective way to ensure that any clinical task intended for a medical image can be performed reliably as the use of poor quality images for clinical tasks can result in inaccurate, or potentially erroneous, diagnoses or measurements \citep{qa_retina, qa_fetal, qa_review}. IQA serves as a mechanism to ensure that intended downstream target tasks for medical images, such as diagnostic, therapeutic or navigational tasks, can be performed effectively and reliably.

The use of IQA in medicine to ensure reliability in clinical task performance is common and various approaches to IQA have been proposed in the past decades to address this problem. Existing IQA methods broadly fall into two categories, manual and automated assessment. Manual assessment is widely used in clinical practice \citep{qa_review}, often involving human interpretation of a set of criteria in order to assess image quality \citep{qa_subj_single_2, qa_subj_single_3, qa_subj_single_4}. Due to the high variance in manual assessment, consensus or mean quality scores from multiple observers may be used to assess images \citep{qa_subj_multiple_1, qa_subj_multiple_2}. Although consensus-based methods are able to reduce the variance in predictions and produce repeatable measurements for IQA, the human cost associated with obtaining IQA scoring from multiple expert observers is high \citep{qa_review}. Automated assessment methods for IQA provide a means to ensure reproducible measurements and to reduce both the variance in predictions and the involvement of human perception of both the medical image and the IQA criteria \citep{qa_review}. These automated methods differ from manual methods in that after a computational tool construction phase, such as training a machine learning model, they are able to quantify IQA without requiring human judgement for any new images being assessed. It is, however, important to note that model construction itself may require human judgement, possibly in the form of labelled samples or to select common features across high or low quality images, for both development and validation of the model.

Automated methods can further be classified based on the extent to which they utilise information from a reference set of images in the IQA model construction phase \citep{qa_review}. Full-reference automated methods use selected reference images directly in order to compute an IQA metric. The metric is often based on a similarity measure between the image being assessed and the subjectively selected reference standard good-quality image. The selection of the reference standard and construction of the metric may be considered as part of model construction \citep{qa_ref_sim_1, qa_ref_sim_2, qa_ref_sim_3, qa_ref_sim_4, qa_ref_sim_5, qa_ref_sim_6, qa_ref_sim_7, qa_ref_sim_8, qa_ref_sim_9, qa_ref_sim_10, qa_ref_sim_11, qa_ref_sim_12, qa_ref_sim_13, qa_ref_sim_14, qa_ref_sim_15, qa_ref_sim_16, qa_ref_sim_17}. These full-reference methods and other reduced-reference methods, which use partial information from a selected reference image set \citep{qa_review}, can thus produce automated, reliable and repeatable measurements. No-reference methods aim to eliminate the subjective selection of a reference standard and do not rely on a reference image set for model construction \citep{qa_no_ref_1, qa_no_ref_3, qa_no_ref_4, qa_no_ref_7, qa_no_ref_task_specific_1, qa_no_ref_task_specific_2, qa_retina, qa_vessel, qa_subj_single_2}. These methods require robust mathematical models to capture common features across low quality images or the statistics of high quality image generation, and are therefore specific to certain modalities or applications. For example, methods have been proposed for MR images \citep{qa_no_ref_4, qa_no_ref_7}, single photon emission CT \citep{qa_no_ref_3},  and CT \citep{qa_no_ref_task_specific_1, qa_no_ref_task_specific_2}. Constructing these modality- or application- specific models may not require human labels of IQA at model construction or inference, however, it requires in-depth knowledge of the noise sources within the imaging modality as well as of the image acquisition process and that of intended use of the images \citep{qa_review}. Learning-based approaches use machine learning methods, including recent deep learning, to automate IQA and provide fast inference. Most of the existing methods learn from subjective expert labels of IQA \citep{qa_fetal, qa_retina_deep, qa_liver_deep, qa_zachary_lung, liao_qa_ec, abdi_qa_ec, lin_qa_multitask, qa_prostate_tpus}, based on a set of pre-selected reference images to varying extent. Most automated IQA methods, including full-, reduced- and no- reference methods, both learning- and non-learning based, were validated against human labels of IQA regardless of the extent to which they use human labels for model construction \citep{qa_review}.

In this work we turn our attention to task-specific IQA methods, when the goal of an IQA method is to ensure that specific downstream tasks can be reliably performed, and task-agnostic IQA methods may not always be effective, efficient or feasible. We propose to use the term `task amenability' to define the usefulness of an image for a specific downstream target task. Most IQA methods, regardless of the extent to which they utilise a reference standard, are often designed to be specific to a certain imaging modality or to a specific anatomy, rather than directly quantify the impact on the downstream clinical task performance \citep{qa_review, qa_fetal, qa_liver_deep, qa_no_ref_task_specific_1, qa_no_ref_task_specific_2}, as discussed above. For example, a strong ultrasound reflection, obstructing a significant part of gland boundaries, is catastrophic to a downstream task of gland segmentation, but may not adversely impact a classification task for identifying the presence of the gland. Similar examples also include severe noise found outside of regions of interest, upon which the diagnosis decision does not rely. 

Perhaps more interestingly, as increasing number of downstream tasks are being automated by, for example, machine learning models, the subjectively perceived task-specific IQA may be very different from the actual impact of an image on the automated target task. Thus the extent to which these human-defined criterion can measure impact on machine-automated task performance is still an open question. Thus one could argue that, where the downstream target task is independently learned or performed, task amenability may be difficult to quantify.

\subsection{Related work and contribution}

In our previous work \citep{saeed_amenability}, we proposed a method to objectively quantify task amenability for a specific task by jointly learning the task-specific IQA and the target task to capture the inter-dependence between the two functions, without the need for human labels of IQA. We proposed to simultaneously learn a task predictor, which performs a downstream target task, and an IQA controller, which selects or weights images based on their task amenability. In this scenario, images can be selected or weighted based on their task amenability such that this selected or weighted subset results in improved target task performance. It should be noted that only performing the target task (i.e. using a task predictor function with fixed weights) while learning the task amenability may also be able to capture the dependence of the controller on the target task. However, to capture the inter-dependence which arises from the task predictor training data modification by the controller and the controller training based on task predictor performance, requires the target task to be learnt within the framework. Nonetheless, the fixed task predictor formulation may be useful for applications where the task-predictor has been pre-trained with different data or in applications where what the IQA entails does not impact task performance.

In the proposed problem setting, optimising the controller is dependent on the task predictor being optimised. The problem can thus be modelled in a meta-learning framework where the downstream target task performance is maximised with respect to the controller selected images. The task predictor and the data used to train it are considered to be contained within an environment which reflects a Markov decision process (MDP).

Meta-learning problems have increasingly been formulated as reinforcement learning (RL) problems under RL-based meta-learning. In the RL-based meta-learning framework, a parameter associated with the target task is modified by the controller such that the target task performance can be maximised. The reward, which indicates how well the target task is performed, is computed after the parameter modification and thus indicates the effect of the controller's modification. This reward is used to update the controller in a way which maximises the cumulative obtained reward and thus allows for a parameter setting that maximises the target task performance, to be learnt. The target task can include any automated classification, regression or segmentation task and the parameter modification by the controller can include selecting a data transformation strategy for augmentation policy search \citep{autoaugment, adv_autoaugment}, selecting hyper-parameters like filters for convolutional layers for neural architecture search \citep{archisearch} and sampling training data for data valuation \citep{google_dvrl}. This approach to learn task amenability bears some resemblance with the data valuation framework presented in \citep{google_dvrl}, however, there are several differences in reward formulation without human labels of quality, the RL algorithms used, and other methodological details including use of the controller for holdout data.

In another previous work we proposed a meta-RL training scheme for the controller \citep{saeed_meta_rl}. The training scheme involved a recurrent neural network (RNN) based controller and sampling different MDP environments, each with different observer labels for the target task, during training in order to equip the trained controller with adaptability to new labelling standards. The resulting adaptability from data labelled by non-expert observers to high-quality expert labelled data, carefully curated by reviewed consensus, proved useful for the efficient use of labelled data.

In this work, we summarise the two preliminary sets of experimental results \citep{saeed_amenability, saeed_meta_rl} and present a general framework to learn an adaptable task amenability assessment using meta-reinforcement learning (meta-RL). The proposed scheme samples environments for training, from a distribution of MDP environments, such that the controller can adapt to new environments sampled from the distribution with a few interactions. Adaptive behaviour is learnt as a result of the proposed training scheme which involves sampling multiple environments and the use of a recurrent neural network (RNN) to equip adaptability to the controller similar to the framework proposed in our previous work \citep{saeed_meta_rl}. However, different from the previous work, the distribution of environments in the general framework can be over different target tasks, imaging modalities, observer labels, task predictor functions (e.g. different network architectures), or any other variable within the environment, as opposed to over different observer labels (where each sampled environment can be considered a new meta-task). Meta-RL allows for the task predictor and controller functions to be trained together to capture the inter-dependence between them and also to equip adaptability to the controller over different meta-tasks. Therefore, the focus of this work is to formulate a general meta-RL framework for IQA which is applicable to a variety of scenarios where adaptability is to be learnt over a distribution of environments or meta-tasks.

Contributions of this work are summarised as follows: 1) we present a literature review with a contextual discussion for positioning the proposed task-specific IQA methodology and explains the need for co-learning between the IQA controller and a task predictor; 2) building on our preliminary work presented in two recent conference papers~\citep{saeed_amenability, saeed_meta_rl}, we formulate a general, unified meta-RL based meta learning framework to train an adaptable IQA system which directly accounts for the inter-dependence between the task-specific IQA and the target task; 3) we present new experiments to evaluate the hyper-parameters and design choices in our proposed framework; 4) in addition to summarising the two previously reported target tasks of prostate classification and segmentation, we also evaluate for a new diagnostic target task of pneumonia detection, on a public dataset of chest X-ray images; 5) the code used in this study are made available for reproducibility of the presented experimental results.

\section{Methods}

    \begin{figure*}[!ht]
    \centering
    \includegraphics[width=\textwidth]{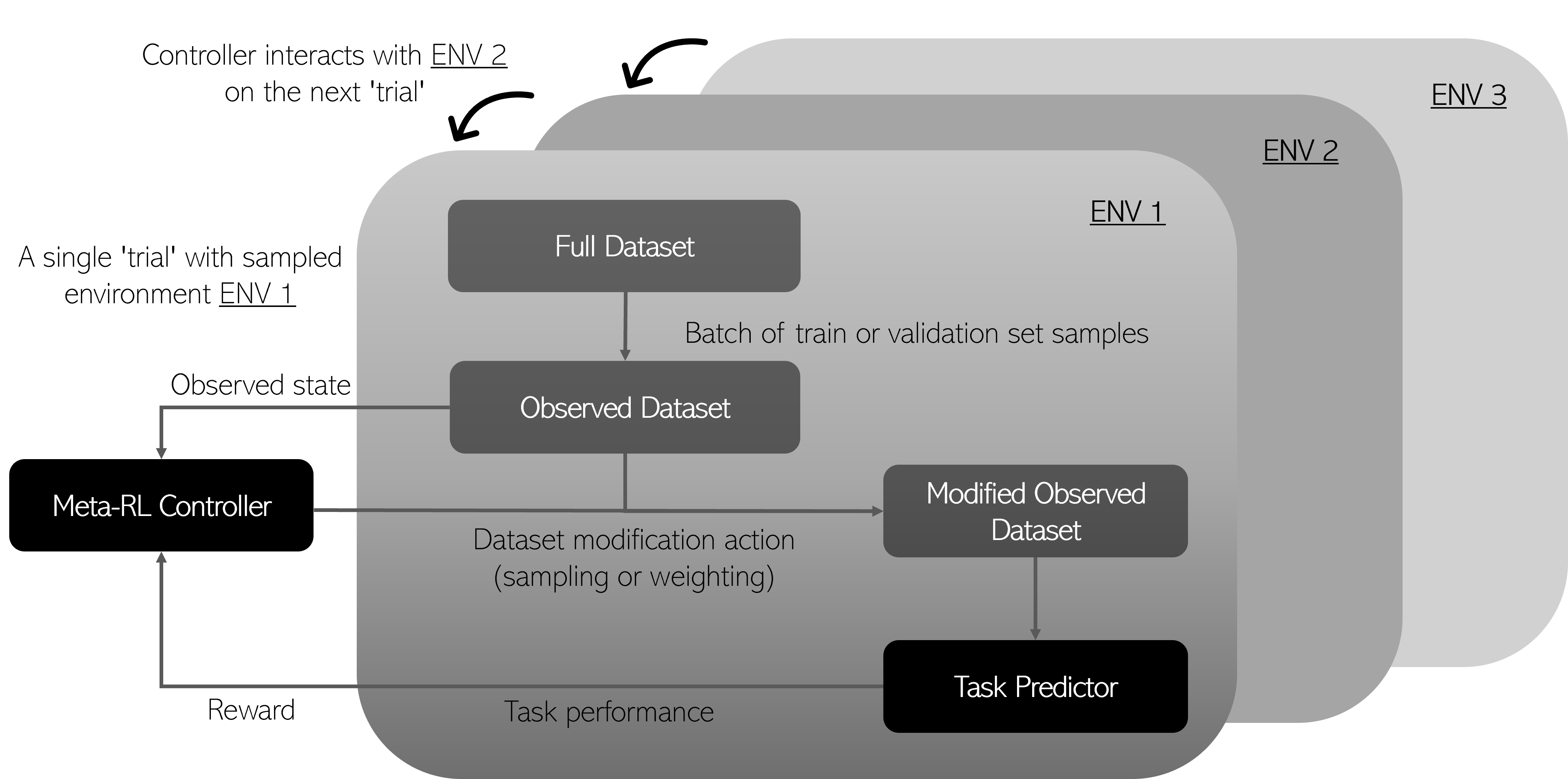}
    \caption{Illustration of the proposed multi-environment meta-RL task amenability framework.}
    \label{fig:meta_iqa}
    \end{figure*}
    
    \subsection{Problem formulation}
    In this section, we formulate the image quality assessment problem in a specific scenario, that uses the output of the measure of quality for task amenable data selection. As described in Sect. \ref{sec:introduction} and illustrated in the Fig. \ref{fig:meta_iqa}, the proposed IQA formulation relies on two inter-dependent functions, the controller and the task predictor.

    \subsubsection{Task predictor and IQA controller}
    
    Without loss of generality, assume the task predictor is a parametric function, 
    \begin{equation}\label{eq:task_pred}
    f(x; w): \mathcal{X} \rightarrow \mathcal{Y}, 
    \end{equation}
    which performs the target task given an image sample $x\in\mathcal{X}$ and outputs a prediction $y\in\mathcal{Y}$, with parameters $w$. 
    The controller is also a parametric function, 
    \begin{equation}\label{eq:controller}
    h(x; \theta): \mathcal{X} \rightarrow[0,1], 
    \end{equation}
    which outputs a task amenability score given an image sample, $x$, with parameters $\theta$. In this formulation $\mathcal{X}$ and $\mathcal{Y}$ correspond to the image and label domains for the specific downstream target task respectively. $\mathcal{P}_X$ and $\mathcal{P}_{XY}$ denote the image and joint image-label distributions with probability density functions $p(x)$ and $p(x,y)$, respectively.
    
    The task predictor performs the downstream target task and the controller selects or weights data used for training the task predictor. The task performance informs the controller decisions over time in order to allow for task performance to be improved. The training methodology outlined below helps to capture this inter-dependence between the two functions.
    
    \subsubsection{Optimising task predictor}
    
    Given that a loss function, $L_f: \mathcal{Y} \times \mathcal{Y} \rightarrow \mathbb{R}_{\geq0}$, measures how well the target task is performed by the task predictor $f(x; w)$ given task label $y$, the task predictor is optimised by minimising a weighted loss function as follows:
    
    \begin{equation}\label{eq:task_pred_opt}
    \min_w \mathbb{E}_{(x,y)\sim\mathcal{P}_{XY}}[L_f(f(x;w), y)h(x;\theta)].
    \end{equation}
    
    Here, weighting by the controller-measured task amenability for the same image sample, $x$, ensures that high loss samples with low task amenability should be weighted less. This provides an incentive to reject such samples, and to accept samples with high task amenability, in optimising the controller described as follows. 
    
    \subsubsection{Optimising controller}
    
    The controller is optimised by minimising a weighted metric function on the validation set $L_h: \mathcal{Y} \times \mathcal{Y} \rightarrow \mathbb{R}_{\geq0}$:
    
    \begin{align}\label{eq:controller_opt}
    \min_\theta \mathbb{E}_{(x,y)\sim\mathcal{P}_{XY}}[L_h(f(x;w), y)h(x;\theta)],\\
    \text{s.t.}\quad \mathbb{E}_{x\sim\mathcal{P}_{X}}[h(x;\theta)] \geq c > 0.
    \end{align}
    
    The controller thus predicts lower task amenability for samples with higher values from this metric function, which translates to lower task performance, due to the weighted sum being minimised. Intuitively, this means correctly predicting the task labels for lower task amenability samples tends to be difficult. The trivial solution of $h\equiv0$ is prevented by introducing the constraint. 
    
    \subsubsection{Bi-level optimisation for learning task amenability}
    
    The proposed task amenability framework can thus be posed as the following bi-level minimisation problem \citep{sinha_bilevel}:
    
    \begin{align}\label{eq:min_prob}
    && \min_\theta \mathbb{E}_{(x,y)\sim\mathcal{P}_{XY}}[L_h(f(x;w^*), y)h(x;\theta)],\\
    \text{s.t.}&& w^*=\arg\min_w \mathbb{E}_{(x,y)\sim\mathcal{P}_{XY}}[L_f(f(x;w), y)h(x;\theta)],\\
    &&\mathbb{E}_{x\sim\mathcal{P}_{X}}[h(x;\theta)] \geq c > 0.
    \end{align}
    
    This problem can be re-structured to permit sampling or selection based on controller outputs by considering the data $x$ and $(x,y)$ to be sampled from the controller-selected or -sampled distributions $\mathcal{P}_{X}^h$ and $\mathcal{P}_{XY}^h$, with probability density functions $p^h(x) \propto p(x)h(x;\theta)$ and $p^h(x,y) \propto p(x,y)h(x;\theta)$, respectively. Thus, re-formulating to facilitate sampling or selection, we can re-write the bi-level minimisation problem as follows:
    
    \begin{align}\label{eq:min_prob_sampling}
    && \min_\theta \mathbb{E}_{(x,y)\sim\mathcal{P}_{XY}^h}[L_h(f(x;w^*), y)],\\
    \text{s.t.}&& w^*=\arg\min_w \mathbb{E}_{(x,y)\sim\mathcal{P}_{XY}^h}[L_f(f(x;w), y)],\\
    &&\mathbb{E}_{x\sim\mathcal{P}_{X}^h}[1] \geq c > 0.
    \end{align}

    \subsection{Meta-reinforcement-learning for task amenability}\label{sec:methods_iqa}
    
    The formulated task amenability assessment problem, eq. \ref{eq:task_pred} to \ref{eq:min_prob_sampling}, can be learnt in a RL-based meta-learning framework as formulated in our previous work \citep{saeed_amenability}. In this work we outline a general meta-RL based meta-learning framework to learn adaptable task amenability assessment.
    
        \subsubsection{Markov decision process environment construction}\label{methds:meta_iqa}
        
\begin{figure}
            \centering
            \includegraphics[width=0.495\textwidth]{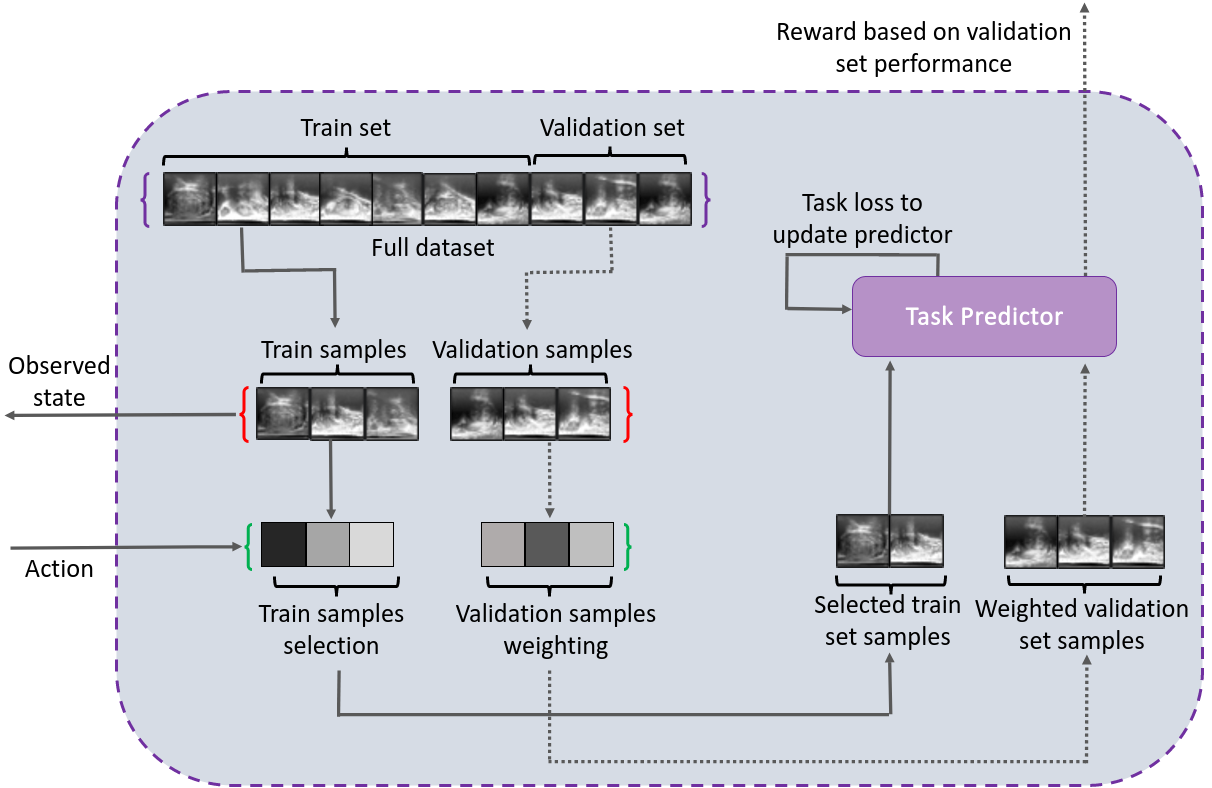}
            \caption{A single environment in the IQA framework.}
            \label{fig:single_env}
        \end{figure}

        The proposed formulation can be modelled as a finite-horizon Markov decision process (MDP) with the controller interacting with, and influencing, an `environment', which contains the task predictor and the data used to train such a function, as illustrated in Fig. \ref{fig:single_env}. The MDP environment for this task amenability problem thus consists of the data from $\mathcal{P}_{XY}$, where this joint image-label distribution is defined as $\mathcal{P}_{XY}= \mathcal{P}_X \mathcal{P}_{Y|X}$, and the target task predictor $f(\cdot;w)$. At time-step $t$, the observed state of the environment $s_t = (f(\cdot; w_t), \mathcal{B}_t)$ is composed of the target task predictor $f(\cdot; w)$ and a batch of samples $\mathcal{B}_t = \{ (x_i, y_i) \}_{i=1}^B$ from a train set $\mathcal{D}_{\text{train}} = \{ (x_i, y_i) \}_{i=1}^N$ from the distribution $\mathcal{P}_{XY}$. If each MDP environment is defined as $M_k$, the joint image-label distribution and task predictor within the environment can be defined as $\mathcal{P}_{XY, k}$ and $f_{k}(\cdot;w_{k})$, respectively. However, in further analysis, we omit $k$ from these expressions, for notational convenience.
        
        \subsubsection{Reinforcement learning for bi-level optimisation}\label{sec:methods_meta_iqa}
    
        Reinforcement learning allows for the training of a controller to maximise a reward obtained based on controller-environment interactions, which are considered to be a MDP. In RL, the MDP is considered to be a 5-tuple $(\mathcal{S}, \mathcal{A}, p, r, \pi)$. $\mathcal{S}$ is the state space and $\mathcal{A}$ is the continuous action space. $p: \mathcal{S} \times \mathcal{S} \times \mathcal{A} \rightarrow [0, 1]$ is the state transition distribution conditioned on state-actions, e.g. $p(s_{t+1} | s_t, a_t)$ represents the probability of the next state $s_{t+1} \in \mathcal{S}$ given the current state $s_t \in \mathcal{S}$ and action $a_t \in \mathcal{A}$. 
        
        The reward function is denoted by $r: \mathcal{S} \times \mathcal{A} \rightarrow \mathbb{R}$ and $R_t = r(s_t, a_t)$ denotes the reward given current state $s_t$ and action $a_t$. The policy, $\pi(a_t|s_t): \mathcal{S} \times \mathcal{A} \in [0, 1]$, represents the probability of performing the action $a_t$ given the state $s_t$. The controller interacting with an environment creates a trajectory of states, actions and rewards, $(s_1, a_1, R_1, s_2, a_2, R_2, \ldots, s_T, a_T, R_T)$, where the subscript indicates the time-step. 
        
        The goal of the agent is to maximise the cumulative reward over a trajectory. The cumulative reward is the discounted sum of accumulated rewards starting from time-step $t$: $Q^{\pi}(s_t, a_t) = \sum_{k=0}^{T} \gamma^k R_{t+k}$, where the discount factor $\gamma\in[0, 1]$ is used to discount future rewards. The objective of the controller is thus to learn a parameterised policy $\pi_{\theta}$ which maximises the expected return $J(\theta)=\mathbb{E}_{\pi_{\theta}}\left[ Q^{\pi}(s_t, a_t) \right]$. The central optimisation problem in RL can be expressed as: 
        
        $$\theta^* = \text{argmax}_{\theta} J(\theta)$$
        where $\theta^*$ denotes optimal policy parameters.
        
        In our previous work \citep{saeed_amenability}, we proposed to train the task-amenability-predicting controller using RL, where the controller outputs sampling probabilities $\{h(x_{i,t}, \theta)\}_{i=1}^B$ based on the input images. The action $a_t = \{a_{i,t}\}_{i=1}^B \in \{0, 1\}^B$ leads to a sample selection decision for target task predictor training, if $a_{i,t}=1$. The selection is done based on $a_{i,t}\sim\text{Bernoulli}(h(x_{i,t}; \theta))$. The policy $\pi_{\theta}(a_t|s_t)$ is defined as:
        
        \begin{align}
        \log \pi_{\theta}(a_t|s_t) = \sum_{i=1}^B h(x_{i,t}; \theta) a_{i,t} + (1-h(x_{i,t}; \theta)(1-a_{i,t}))
        \end{align}
        
        In this formulation, the reward $R_t$ was formulated based on the metric function which measures performance of the target task, $L_h$. There are several options for how to design a metric using $L_h$ and we describe a few examples in the subsequent paragraphs. 
        
        \subsubsection{Meta reinforcement learning for adaptability} 
        
        In this work, we propose to train the controller using meta-RL that unifies the single-environment formulation proposed in \citep{saeed_amenability}. 
        Meta-RL is a training procedure which has the same goal of maximising the expected return as RL, however, the objective is averaged across multiple MDP environments in a distribution of MDPs such that the trained controller can effectively generalise or adapt to new MDPs sampled from the distribution \citep{duan_metarl, wang_metarl, botvinick_metarl}. To facilitate adaptability in the controller, it is shared across different MDPs sampled from the distribution of MDPs $\mathcal{P}_M$. A period of interaction with a single MDP is referred to as a \textit{trial}. The controller learns across multiple trials by sampling an MDP $M_k\sim\mathcal{P}_M$ for each trial. The controller also takes the action $a_t$, raw reward $r_t$, and termination flag $d_t$ at the previous time step in addition to the observed current state $s_{t+1}$. Note that for per-sample operation $r_t=R_t$ at the episode end, and zero otherwise, similar to sparse reward formulations in \citep{duan_metarl, wang_metarl}. Additionally, the controller embeds a recurrent neural network (RNN) with the internal memory shared across episodes within the same trial. The internal memory is reset before the controller encounters another environment i.e. at the start of each trial. This mechanism allows for adaptability even with fixed wights \citep{duan_metarl, wang_metarl, botvinick_metarl}. The additional inputs along with the embedded RNN and its internal memory make the controller a function of the history leading up to a sample such that changing history can influence the action for that sample. It should be noted that each sampled MDP $M_k$, has its own task predictor and joint image-label distributions. There may be benefit in sharing components between the MDPs and a few such cases are discussed in the following paragraphs. The training scheme is outlined in Algorithm \ref{algo:meta_iqa} and summarised in Fig. \ref{fig:meta_iqa}.
        
        Subsequent to training using the meta-RL scheme, adaptation can be carried out by sampling a single MDP $M_a\sim\mathcal{P}_M$, where $M_a$ is the environment to be adapted to, resetting the RNN internal state once at the start of the adaptation and allowing for controller-environment interaction across multiple episodes. This means following the same scheme as Algorithm \ref{algo:meta_iqa} but without sampling a new MDP on each trial, without resetting the RNN internal state at the start of each trial, and without updating the controller. The adaptability in this adaptation scheme is a result of updating RNN internal state rather than updates of the weights.
        
        This meta-RL training and adaptation scheme could potentially be applicable to a wide range of scenarios. Training an IQA system which could adapt to different target tasks, new labels for the target task, different task predictor functions, different reward metrics, or different imaging modalities are all possible with the proposed training scheme. This could allow for efficient use of data without the need to retrain an entirely new IQA system. As an example, to train an IQA system adaptable to different task predictor functions, for example neural network architectures, the defined MDP distribution $\mathcal{P}_M$ would have MDPs each with a task predictor with a different architecture. For some of these applications, such as to train an adaptable IQA system across multiple reward metrics or labels for the target task, it may be beneficial to share the task predictor between MDP environments. This equips adaptability to the task predictor in addition to the controller. Moreover, sharing certain components of the dataset may be useful as well, such as for multiple observer labels, it may be useful to share images across environments and define the distribution of MDPs $\mathcal{P}_M$ to be over different joint image-label distributions as outlined in Sect. \ref{sec:methods_multi_observer}.
        
        \subsubsection{Reward formulation}\label{sec:methods_reward}
        
        In the previous sections we outlined how a performance metric can be used as the reward function to train the task amenability-predicting controller. This allows for many different combinations and variations in the reward formulation, however, in this work we focus on a performance metric computed on the validation set $\mathcal{D}_{\text{val}} = \{(x_j, y_j)\}_{j=1}^M$, from the same distribution as the train set $\mathcal{P}_{XY}$, as: $\{l_{j,t}\}_{j=1}^M = \{L_h(f(x_j;w_t), y_j)\}$. This performance can be used to formulate the un-clipped reward $\tilde{R}_t$. $\tilde{R}_t$ can be formulated in several ways, with or without the controller output for the validation set $\{h_j\}_{j=1}^M = \{h(x_j;\theta)\}_{j=1}^M$. In this work we consider three formulations for $\tilde{R}_t$:
        
        \begin{enumerate}
        \item $\tilde{R}_{\text{avg},t}=-\frac{1}{M}\sum_{j=1}^Ml_{j,t}$, the average performance.
        \item $\tilde{R}_{\text{w},t}=-\frac{1}{M}\sum_{j=1}^Ml_{j,t}h_j$, the weighted sum.
        \item $\tilde{R}_{\text{sel},t}=-\frac{1}{M'}\sum_{j'=1}^{M'}l_{j',t}$, the average of the selected $M'$ samples.
        \end{enumerate}
        
        The first reward formulation $\tilde{R}_{\text{avg},t}$ requires pre-selection of highly task amenable data by a human observer to form the validation set since no controller weighting or sampling is performed. These ``task amenability'' labels can be acquired in addition to the task labels and can be used to form such a clean fixed validation set. The second $\tilde{R}_{\text{w},t}$ and third $\tilde{R}_{\text{sel},t}$ reward formulations do not require these human labels of task amenability as they utilise the controller output to weight or sample the validation set, respectively. For the third reward formulation, $\tilde{R}_{\text{sel},t}$, $\{{j'}\}_{j'=1}^{M'} \subseteq \{j\}_{j=1}^M$ and $h_{j'}\leq h_{k'}, \forall{k' \in\{j'\}^c}, \forall{j'\in\{j'\}}$, i.e. the un-clipped reward $\tilde{R}_{\text{sel},t}$ is the average of $\{l_{j'}\}$ from the subset of $M'=\lfloor (1-s_{rej})M \rfloor$ samples, by removing $s_{rej}\times100\%$ samples from the end, after sorting $h_j$ in decreasing order. During training, this $\tilde{R_t}$ value is clipped using a moving average $\bar{R}_t ={\alpha}_R\bar{R}_{t-1}+(1-{\alpha}_R)\tilde{R}_t$, where ${\alpha}_R$ is a hyper-parameter set to 0.9. It is possible to clip the reward using other values such as using a random selection baseline rather than using a moving average. It is interesting to note that since the validation set is formed of multiple samples, this means that the controller will not only keep samples on which the target task can be performed perfectly but will try to select samples to facilitate generalisability to new samples. Additionally, since the validation set is either weighted or selective, with or without human labels of IQA, with respect to task amenability, generalisability to samples that have high task amenability is encouraged compared to those with low task amenability. This means that cases such as all samples being selected or all but one samples being rejected, are discouraged in this formulation.
    
        \subsection{Single environment cases}\label{sec:methods_single_env}
        
        The task-specific IQA system presented in \citep{saeed_amenability} can be considered a special case of the general meta-RL framework presented in this work. In the single environment case, there are no environment-specific trials but rather we can consider the same environment to be sampled at the start of each trial. Moreover, adaptation is not required after training. The Reptile update (introduced in Sect. \ref{sec:methods_multi_observer}) is also not applicable to this single environment case and the RNN may be replaced with a simple feed-forward function such as a deep neural network and the additional inputs of $a_t$, $r_t$, and $d_t$ may be omitted. More succinctly, meta-RL simplifies to RL for this single environment case. This is useful when adaptability over a dataset distribution is not required but generalisability to new samples in a dataset is still desirable.
    
        \subsection{Multiple environments for inter-observer labels}\label{sec:methods_multi_observer}
        
        Our previous work presented preliminary results for such a specific formulation of the meta-RL framework~\citep{saeed_meta_rl}, in which a trained IQA system could adapt across multiple observers of task amenability, for example to use non-expert target task labels for training and then use limited expert-labelled data for adaptation to adapt the task-specific IQA definition to a new reference standard. In this paper, the general framework presented in Sect. \ref{sec:methods_meta_iqa} is a more general case of the multi-observer framework presented in \citep{saeed_meta_rl}. Additional results from a comprehensive set of experiments are also presented in Sect.~\ref{sec:results}. 
        
        For the multi-observer setting we consider multiple label distributions $\{\mathcal{P}_{Y|X}^{k}\}_{k=1}^K$ which means that each sample $x$ has multiple labels $\{y^k\}_{k=1}^K$. Therefore, we have multiple joint image-label distributions $\mathcal{P}_{XY}^{k}=\mathcal{P}_{X}\mathcal{P}_{Y|X}^{k}$ for $k=1,\ldots,K$. We use each joint image-label distribution to form an individual MDP environment $M_k$. The distribution of MDPs $\mathcal{P}_M$ is thus over multiple observers for the same target task. In the multi-observer framework, weights of the task predictor are synced between the different environments at the start of a new trial and the task predictor is updated using Reptile \citep{nichol_reptile} to allow for adaptability and data efficiency of the predictor in addition to the controller. 
        
        The Reptile-based task predictor update therefore consists of two steps: 1) perform gradient descent for the task predictor $f(\cdot; w_t)$, starting with weights $w_t$ and ending in weights $w_{t, \text{new}}$; 2) update the task predictor weights $w_t \leftarrow w_t + \epsilon(w_{t, \text{new}}-w_t)$. $\epsilon$ is set as 1.0 initially and linearly annealed to 0.0 as trials progress (similar to \citet{nichol_reptile}). Adaptive moment estimation \citep{kingma_adam} is used as the gradient descent algorithm for this work.

    \subsection{Controller selection at inference}
    
    A separate holdout set is used to evaluate the controller-learned task amenability assessment. We remove a proportion of the samples valued least by the controller by sorting the samples in the holdout set according to controller predicted values. This ratio of samples removed based on controller outputs is referred to as the `holdout set rejection ratio'.

\begin{algorithm}[!ht]
\SetAlgoLined
\KwData{Multiple MDPs $M_k\sim\mathcal{P}_M$.}
\KwResult{Controller $h(\cdot;\theta)$.}
\BlankLine
\While{not converged}{
Sample an MDP $M_k\sim\mathcal{P}_M$\;
Reset the internal state of controller $h$\;
    \For{Each episode in all episodes}{
        \For{$t\leftarrow 1$ \KwTo $T$}{
            Sample a training mini-batch $\mathcal{B}_{t}=\{(x_{i,t},y_{i,t})\}_{i=1}^{B}$\;
            Compute selection probabilities $\{h_{i,t}\}_{i=1}^B=\{h(\tau_{i,t};\theta_t)\}_{i=1}^B$\;
            Sample actions $a_{t}=\{a_{i,t}\}_{i=1}^B$ w.r.t. $a_{i,t}\sim\text{Bernoulli}(h_{i,t})$\;
            Select samples $\mathcal{B}_{t,\text{selected}}$ from $\mathcal{B}_{t}$\;
            Update predictor $f(\cdot;w_t)$\ with $\mathcal{B}_{t,\text{selected}}$\;
            Compute reward $R_{t}$\;
        }
        Collect one episode $\{\mathcal{B}_t,a_t,R_t\}_{t=1}^T$\;

        Update controller $h(\cdot;\theta)$ using the RL algorithm;
    }
}
\caption{Adaptable image task amenability assessment using multiple environments}
\label{algo:meta_iqa}
\end{algorithm}

\section{Experiments}
    
We conduct experiments to evaluate the efficacy of the proposed task amenability framework, both in single and multiple environment settings. The details of the experiment and the data used to to evaluate the proposed framework are outlined in this section.
    
    \subsection{Experiment data}
        
        We used two datasets in our experiments; the first dataset was formed of trans-rectal ultrasound (TRUS) images of the prostate gland and surrounding regions, and the second dataset consisted of chest X-ray images. The TRUS dataset allows us to evaluate the efficacy of the proposed framework to two common surgical guidance and navigational target tasks of prostate presence detection and gland segmentation. The chest X-ray dataset allows us to evaluate the framework for a common diagnostic target task of pneumonia detection using chest X-ray images, and to demonstrate that the proposed approach is not limited to a single imaging modality, dataset, and target task. Moreover, this dataset is publicly available and commonly used for medical imaging research.
        
        \subsubsection{Transrectal ultrasound imaging and the target tasks}
        
        Ultrasound guided biopsy procedures were performed for $259$ patients as part of the clinical trials (NCT02290561, NCT02341677). Trans-rectal ultrasound images were acquired for these patients  using a side firing transducer of a bi-plane trans-perineal ultrasound probe (C41L47RP, HI-VISION Preirus, Hitachi Medical Systems Europe). These images were acquired either while manually positioning a digital trans-perineal stepper (D\&K Technologies GmbH, Barum, Germany) for navigation using ultrasound or while rotating the stepper with recorded relative angles to scan the entire gland. Each image consisted of 50-120 2D frames of TRUS with relative angles recorded for each frame.
        
            \paragraph{Data pre-processing}
            
            In order to feasibly label the acquired data, the resulting TRUS images were sampled at approximately $4$ degrees resulting in a total of $6712$ 2D frames of TRUS from $259$ patients. These images were randomly split into train, validation and holdout sets, maintaining patient-level separation between sets, with $4689$, $1023$, and $1000$ images from $178$, $43$, and $38$ subjects, respectively. All the images were labelled by four observers for two downstream target tasks (see subsequent paragraph for further details). In addition to the labels for target task, human labels of task amenability for both target tasks were acquired for the purpose of comparing to controller predicted task amenability assessment and in order to select data for the fixed clean validation set reward strategy. These task amenability labels were in the form of binary labels indicating if an image was amenable for the target task or not. 
            
            \paragraph{Two target tasks and their labels}
            
            Labels for two target tasks were curated for all the images: 1) prostate presence classification (binary scalar value indicating prostate presence); 2) prostate gland segmentation (binary image mask of the prostate gland). 
            
            Three sets of labels were collected from three trained medical imaging researchers for both target tasks. For brevity, these three sets are referred to as ``non-expert'' labels and are denoted by $\{L_i\}_{i=1}^3$. An additional set of consensus labels were generated using a slice-level and pixel-level majority vote for the classification and segmentation tasks, respectively. These consensus labels are denoted by $L_C$. These consensus labels were then reviewed by a urologist and edited as required. The labels curated by reviewed consensus with an ``expert'' are denoted as $L_R$. The three trained medical imaging researchers each had between 2-5 years of experience with medical images and their annotation whereas the urologist had over 15 years of experience.
            
            \paragraph{Summary of experiments}
            
            With this dataset we perform two types of experiments: 1) experiments where we use the consensus labels $L_C$ as a single environment for training using the single environment simplified case presented in Sect. \ref{sec:methods_single_env}; 2) experiments where we use the non-expert label sets $\{L_i\}_{i=1}^3$ for training and the expert label set $L_R$ for adaptation using the multi-environment framework (more specifically the multi-observer setting presented in Sect. \ref{sec:methods_multi_observer}). Experiments that fall under type 1 are referred to as single-environment experiments. We investigate the effect of the three proposed reward strategies, presented in Sect. \ref{sec:methods_reward}, and compare with a non-selective baseline (Sect. \ref{sec:exp_rewards}), compare controller-predicted task amenability to human labels of IQA (Sect. \ref{sec:exp_human_iqa}), conduct a sensitivity analysis for the $s_\text{rej}$ hyper-parameter for the segmentation task (Sect. \ref{sec:exp_sens_an}). For experiments that fall under type 2, referred to as multi-environment experiments, we compare the proposed multi-environment meta-RL framework to a single environment baseline (Sect. \ref{sec:exp_multi_env}), and conduct ablation studies to evaluate the design choices of the proposed meta-RL algorithms and its training strategy (Sect. \ref{sec:exp_multi_env}).

        \subsubsection{Public chest X-ray data and the target task}
        
        A public dataset of chest X-ray images with binary labels indicating pneumonia diagnosis \citep{medmnist} was used to evaluate the applicability of the proposed framework to diagnostic target tasks. A total of 5856 images were randomly split into train, validation and holdout sets with 4708, 524 and 624 images, respectively. The images are scaled down paediatric chest X-ray images acquired as part of routine clinical visits. Since low-quality images were manually removed from this data we added artificial corruptions of random intensities to these images in order to evaluate the task amenability assessment framework. Two corruption operations with random intensities were used. Firstly, random Gaussian noise was added to images with intensities varying between 0.0 and 0.8 where an intensity of 1.0 means all pixels in the image are corrupted by random noise and 0.0 means that no pixels are corrupted (percentage of pixels corrupted in the image varies linearly with intensity value). Secondly, random obstructions were added to the image with intensities between 0.0 and 1.0 where an intensity of 1.0 means 100\% of the image is obstructed by zero intensity pixels and 0.0 means that the original image is fully visible with no obstructions added (percentage of image obstructed varies linearly with intensity value). 
        
        The dataset used and the associated experiments are available in an open-source GitHub repository: \url{https://github.com/s-sd/task-amenability}. 
        
        \paragraph{Summary of experiments}
        
        With this dataset we perform experiments with the reduced single environment (Sect. \ref{sec:methods_single_env}) and thus experiments for this data all fall under single-environment experiments. We evaluate different RL algorithms within the proposed framework (Sect. \ref{sec:exp_rl_algo}), and compare the performance of the task predictor when controller selection is applicable to the holdout set with random selection and non-selective baselines (Sect. \ref{sec:exp_comp_rand_non}).

    \subsection{IQA performance evaluation}
    
    The IQA system, including both the controller and target task predictor are evaluated jointly using task performance. This serves as a direct measure of performance for the task predictor and indirect measure of performance for the controller with respect to the learned task amenability. The task performance measures for the different target tasks considered in this work are outlined as follows.
    
        \paragraph{Prostate presence detection} 
        For the prostate presence detection classification task, we report the mean accuracy (Acc.) for the holdout set with standard deviation (St.D.) as a measure of inter-patient variability.
        
        \paragraph{Prostate gland segmentation}
        We report the mean binary Dice score (Dice) on the holdout set with St.D. to measure inter-patient variability, for the prostate gland segmentation task.
        
        \paragraph{Pneumonia detection}
        For the pneumonia detection classification task, we report Acc. and we use bootstrap sampling to estimate the St.D. as a measure of inter-sample variance for the holdout set.
        
    All results, for all tasks, are averaged across all the samples in the holdout set. Where controller selection on the holdout set is applicable, the holdout set rejection ratio is specified and the task performance measures are computed over the selected samples only. Paired t-test results at a significance level of $5\%$ are reported for comparisons.
        
    \subsection{Training details}
    
        \paragraph{Prostate presence detection} 
        For the single-environment class of experiments, the controller is trained using the deep deterministic policy gradient (DDPG) \citep{ddpg} algorithm with hyper-parameters being empirically configured. For the multi-environment type of experiments the controller is trained using proximal policy optimisation (PPO) \citep{schulman_ppo}. An Alex-Net-style architecture \citep{alexnet_class} was used as the target task predictor. The target task predictor and controller were trained with a cross-entropy loss and reward based on classification accuracy (Acc.) for the classification task. Hyper-parameters for the controller and task predictor network architectures and training procedures remain unchanged from default unless specified. The actor and critic networks used in the DDPG and PPO algorithms pass the image inputs via a 3-layered convolutional encoder. These networks then feed into 3 fully connected layers (which embed an RNN in the case of PPO as detailed in Sect. \ref{sec:exp_multi_env}).
        
        \paragraph{Prostate gland segmentation}
        The DDPG and PPO algorithms were used for the single- and multi-environment experiments, respectively, similar to the prostate presence detection task and hyper-parameters for these controller networks are also identical between these two tasks. The U-Net architecture \citep{unet_seg} was used as the task predictor and was trained with a pixel-wise cross-entropy loss. The reward was based on mean binary Dice score.
        
        \paragraph{Pneumonia detection}
        All details including the target task loss, reward and RL hyper-parameters remain the same as the prostate presence classification task, however since there are no multi-environment experiments for this dataset, for comparison between different RL algorithms, both DDPG and PPO are used in the single-environment setting for controller training.

\subsection{Single-environment experiments}
    
    \subsubsection{Evaluating different reward strategies}\label{sec:exp_rewards}
    
    The proposed RL-based task amenability framework for a single environment was evaluated using the TRUS data with consensus labels $L_C$ for training for all three reward strategies presented in Sec \ref{sec:methods_reward}. 
    
    To evaluate the relationship of the controller output with the target task performance, different percentages of the holdout set were removed according to controller output and the mean performance measure and measure of spread are reported for the remaining samples. The three reward formulations were compared with a baseline target task predictor with no controller selection during training or testing. For the fixed clean validation set reward formulation, $s_{\text{rej}}$ was set as 0.05 and 0.15 for the prostate classification and segmentation tasks, respectively, for all experiments, unless specified otherwise. The holdout set rejection ratio for the prostate presence detection and gland segmentation tasks are set to 0.05 and 0.15, respecitvely.
    
        \paragraph{Sensitivity analysis for the rejection ratio for the selective reward formulation}\label{sec:exp_sens_an}
        
        The selective reward formulation has an additional hyper-parameter $s_{\text{rej}}$. We conducted a sensitivity analysis for this hyper-parameter for the segmentation task for the TRUS data. The performance measure and spread are reported for this task for varying $s_{\text{rej}}$ using the selective reward formulation. For the purpose of comparison, the task performance is reported for a holdout set rejection ratio of $0.15$ for all tested values of $s_{\text{rej}}$.
        
        \paragraph{Comparing controller labels to human labels of task amenability}\label{sec:exp_human_iqa}
        
        We compared human labels of task amenability with controller predictions for the two target tasks for the TRUS data. These comparisons were made for all three reward strategies and for the purpose of comparing with binary human-labelled task amenability, $5\%$ and $15\%$ of the lowest valued controller samples were considered as having low-task amenability. These comparisons are presented in the form of contingency tables.

    \subsubsection{Controller selection and a non-selective baseline}\label{sec:exp_comp_rand_non}
    
    The weighted validation set strategy was used to evaluate the efficacy of the proposed RL-based task amenability framework on the chest X-ray dataset, without using any human labels of task amenability. The relationship between target task performance and controller output was studied by removing a proportion of samples based on controller output. The results are reported for a holdout set rejection ratio of $0.10$. Two baseline networks trained and tested, 1) without controller selection referred to as the non-selective baseline, and 2) with a random selection strategy, i.e. a trivial controller, referred to as the random selection baseline, are quantitatively compared.
    
    \subsubsection{Comparing different RL algorithms}\label{sec:exp_rl_algo}
    
    To evaluate the sensitivity of the proposed framework to different RL algorithms. The proximal policy optimisation (PPO) and deep deterministic policy gradient (DDPG) algorithms were compared for controller training, based on the weighted validation set reward strategy with the chest X-ray dataset. The holdout set rejection ratio for this experiment was set as $0.10$.

    \subsection{Multi-environment for multi-observer labelling}\label{sec:exp_multi_env}
    
    To evaluate the multi-environment meta-RL framework, for the multi-observer setting, we trained and compared three models for task amenability:
    
    \begin{itemize}
        \item \textit{Meta-baseline}: This model was trained with all the high-quality reviewed consensus labels $L_R$. Only a single environment was used to train with this ``expert'' labelled data to establish a reference.
        \item \textit{Meta-RL}: The proposed meta-RL framework was used for training with the three non-expert labels $\{L_i\}_{i=1}^3$ forming the three environments for training where the train and validation sets are used. The task predictor and controller were subsequently adapted using $k\times100\%$ of the training and validation sets with expert labels $L_R$.
        \item \textit{Meta-RL Variant}: We conduct an ablation study in order to evaluate the effectiveness of the environment-level separation, we trained a model using all of the training and validation data using non-expert labels $\{L_i\}_{i=1}^3$ as a single environment. That is to say that the trials were not environment specific and Reptile was not used to perform updates to optimise the task predictor as it reduces to gradient descent in this case. The task predictor and controller were adapted using $k\times100\%$ of the training and validation sets with expert labels $L_R$ but for this model, the RNN internal state was not reset before adaptation.
    \end{itemize}
    
    We evaluate these models for varying $k$-values, where $k$ is the ratio of expert-labelled samples used for adaptation ($k\times100\%$ samples used).
    
    The TRUS data was used for this experiment and the multi-environment framework was evaluated for both target tasks. The controller, for this experiment, was embedded with an RNN and had additional inputs of the previous reward, action and terminal flag. The image was passed through 3 convolutional layers before being passed to the RNN embedded controller which had a stacked LSTM architecture with hyper-parameters remaining unchanged from defaults in \citep{wang_metarl}. PPO was used to train the controller in this meta-RL framework since the DDPG algorithm relies on random sampling from a replay buffer which is not well suited when adaptation is done based on RNN internal state updates, which builds sequential memory in the system.

\section{Results}\label{sec:results}




\begin{table*}[!ht]
\centering
\caption{Results on the controller-selected holdout set.}
\begin{tabular}{|c|c|c|}
\hline
Task & Reward computation strategy & Mean $\pm$ St.D.\\
\hline
\multirow{4}{7em}{Prostate presence (Acc.)} & Non-selective baseline & 0.897 $\pm$ 0.010\\
\cline{2-3}
& $\tilde{R}_{\text{avg},t}$, fixed validation set & 0.935 $\pm$ 0.014\\
\cline{2-3}
& $\tilde{R}_{\text{w},t}$, weighted validation set & 0.926 $\pm$ 0.012\\
\cline{2-3}
& $\tilde{R}_{\text{sel},t}$, selective validation set & 0.913 $\pm$ 0.012\\
\hline
\multirow{4}{7em}{Prostate segmentation (Dice)} & Non-selective baseline & 0.815 $\pm$ 0.018\\
\cline{2-3}
& $\tilde{R}_{\text{avg},t}$, fixed validation set & 0.890 $\pm$ 0.017\\
\cline{2-3}
& $\tilde{R}_{\text{w},t}$, weighted validation set & 0.893 $\pm$ 0.018\\
\cline{2-3}
& $\tilde{R}_{\text{sel},t}$, selective validation set & 0.865 $\pm$ 0.014\\
\hline
\multirow{3}{7em}{Pneumonia detection (Acc.)} & Non-selective baseline & 0.817 $\pm$ 0.026\\
\cline{2-3}
& $\tilde{R}_{\text{w},t}$, weighted validation set (DDPG) & 0.843 $\pm$ 0.033\\
\cline{2-3}
& $\tilde{R}_{\text{w},t}$, weighted validation set (PPO) & 0.838 $\pm$ 0.031\\
\hline
\end{tabular}
\label{tab:res_IQA}
\end{table*}

\begin{figure}[!ht]
  \centering
  \subfloat[Prostate presence classification task]{\includegraphics[width=0.45\textwidth]{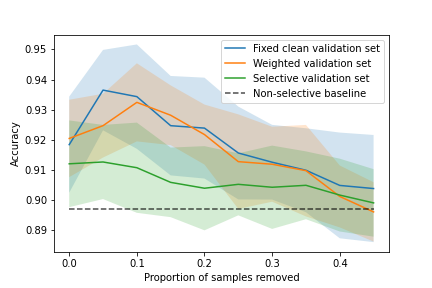}\label{fig:acc_reject}}
  \hfill
  \subfloat[Prostate segmentation task]{\includegraphics[width=0.45\textwidth]{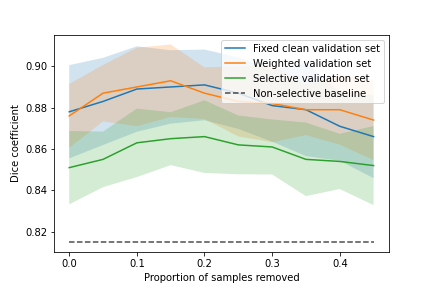}\label{fig:dice_reject}}
  \caption{Plots of the task performance (in respective Acc. and Dice metrics) against the holdout set rejection ratio for the two tasks. (Colour figure)}
  \label{fig:reject_IQA}
\end{figure}

\paragraph{Reward strategies} The results of Acc. and Dice from the TRUS dataset, with the three proposed reward strategies, are summarised in Table \ref{tab:res_IQA}. The performance measures are computed after a controller selection of the holdout set with holdout set rejection ratios of 0.05 and 0.15 for the prostate presence classification and prostate segmentation tasks, respectively. For both of the tasks, all three proposed reward formulations were able to achieve higher performance compared to a non-selective baseline, with statistical significance (\textit{p-values$<$0.001}). The fixed clean validation set and weighted validation set strategies led to significantly higher performance (\textit{p-values$<$0.001}) compared to the selective validation set reward formulation, for both tested tasks. However, significance was not observed between those from the fixed clean validation set and weighted validation set reward strategies, for both the classification (\textit{p-value=0.06}) and segmentation (\textit{p-value=0.49}) tasks. To evaluate performance at varying holdout set rejection ratios, the mean performance against holdout set rejection ratio is plotted in Fig. \ref{fig:reject_IQA}. The peak classification Acc. $0.935$, $0.932$ and $0.913$ occur at $5\%$, $10\%$ and $5\%$ rejection ratios, for the fixed-, weighted- and selective reward formulations, respectively, while the peak segmentation Dice $0.891$, $0.893$ and $0.866$ occur at $20\%$, $15\%$ and $20\%$ rejection ratios, respectively.

\begin{figure*}[!ht]
\centering
\includegraphics[width=0.8\textwidth]{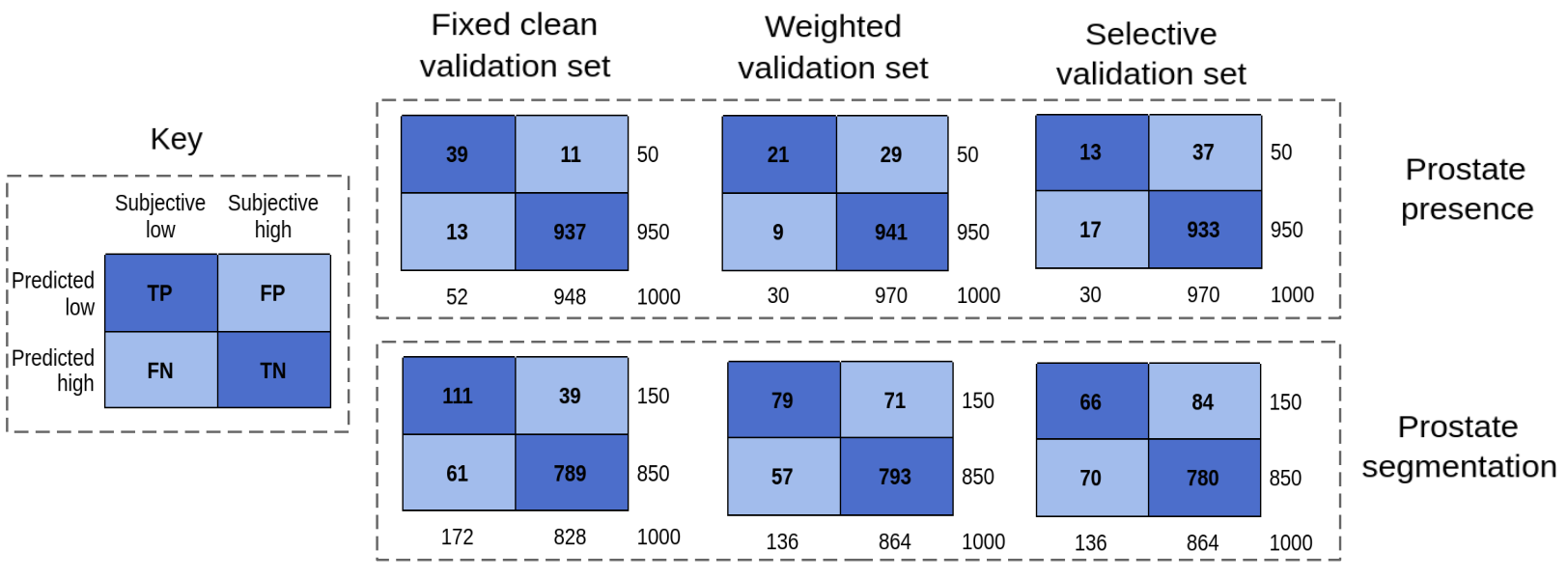}
\caption{Contingency tables comparing subjective labels to controller predictions for the different reward computation strategies. (Colour figure)}
\label{fig:cm}
\end{figure*}

\paragraph{Comparison to human IQA labels} Contingency tables are presented in Fig. \ref{fig:cm} to compare controller predicted task amenability to human labels of task-specific quality for the holdout set, for the prostate presence classification and prostate segmentation tasks. For the purpose of comparison, the holdout set rejection ratio is set to $0.05$ and $0.15$, for the classification and segmentation tasks, respectively, such that rejected samples are considered low task amenability and the rest are considered high task amenability. Agreement in low task amenability samples of $75\%$, $70\%$ and $43\%$, with Cohen's kappa values of $0.75$, $0.51$, was seen for the prostate presence classification task, for the fixed-, weighted- and selective validation sets, respectively. In the segmentation task, agreement on low task amenability samples of $65\%$, $58\%$ and $49\%$, with Cohen's kappa values of $0.63$, $0.48$ and $0.37$, was observed for the three reward formulations, respectively. The accuracy, precision and recall, computed based on these comparisons, for the prostate presence classification task, are: 0.98, 0.78 and 0.75; 0.96, 0.42 and 0.70; and 0.95, 0.26 and 0.43; for the fixed-, weighted- and selective validation set reward strategies, respectively. Similarly for the prostate gland segmentation task, the accuracy, precision and recall are: 0.90, 0.74 and 0.65; 0.87, 0.53 and 0.58; 0.846, 0.44 and 0.49; for the fixed-, weighted- and selective validation set reward strategies, respectively.

\paragraph{Sensitivity analysis and ablation studies} The validation set rejection ratio $s_{\text{rej}}$ is treated as a hyper-parameter and a sensitivity analysis for this hyper-parameter is conducted for the prostate segmentation task. The performance at varying values of $s_{\text{rej}}$ is presented in Table \ref{tab:res_s_rej}. $s_{\text{rej}}$ is increased in increments of $0.05$ and each step increase leads to a statistically significant improvement in performance up to $s_{\text{rej}}=0.20$ (\textit{p-values$<$0.01}). No significance was found comparing performances for a step increase of $s_{\text{rej}}$ from $0.20$ to $0.25$ (\textit{p-value=0.37}). A subsequent step increase from $0.25$ to $0.3$ led to a statistically significant performance reduction (\textit{p-value$<$0.01}).

\begin{table}[!ht]
\centering
\caption{Sensitivity analysis for $s_{\text{rej}}$ for the prostate segmentation task for the selective reward formulation.}
\begin{tabular}{|c|c|}
\hline
$s_{\text{rej}}$ & Mean $\pm$ St.D.\\
\hline
0.00 & 0.827 $\pm$ 0.011\\
\hline
0.05 & 0.838 $\pm$ 0.012\\
\hline
0.10 & 0.845 $\pm$ 0.010\\
\hline
0.15 & 0.865 $\pm$ 0.014\\
\hline
0.20 & 0.882 $\pm$ 0.017\\
\hline
0.25 & 0.888 $\pm$ 0.016\\
\hline
0.30 & 0.876 $\pm$ 0.012\\
\hline
\end{tabular}
\label{tab:res_s_rej}
\end{table}

\begin{figure}[!ht]
    \centering
    \includegraphics[width=0.45\textwidth]{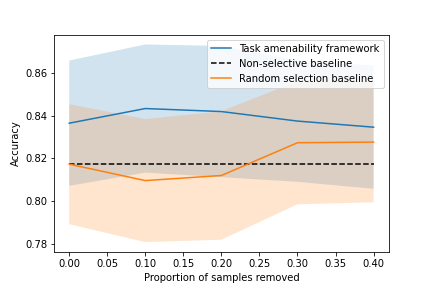}
    \caption{Plot of performance (Acc.) against holdout set rejection ratio for pneumonia detection task. (Colour figure)}
    \label{fig:reject_pneumonia}
\end{figure}

To evaluate the sensitivity of the proposed framework to different RL algorithms, we present task performance results for the pneumonia detection task for a holdout set rejection ratio of $0.10$ in Table \ref{tab:res_IQA}. Although DDPG showed improved performance, significance was not observed (\textit{p-value=0.20}) for the difference between the two algorithms. Both the DDPG- and PPO-trained controller showed significantly improved performance compared with the non-selective baseline (\textit{p-values$<$0.001}). To evaluate the relationship between performance and holdout set rejection ratio, we present a plot of mean performance against holdout set rejection ratio in Fig. \ref{fig:reject_pneumonia}, which presents a comparison between the task amenability framework, a non-selective baseline and a random selection baseline. An implementation of the proposed framework, with both the PPO and DDPG algorithms, along with the data, code, and results for the pneumonia detection task are available in the same open-source GitHub repository.

\begin{figure}[!ht]
    \centering
    \includegraphics[width=0.45\textwidth]{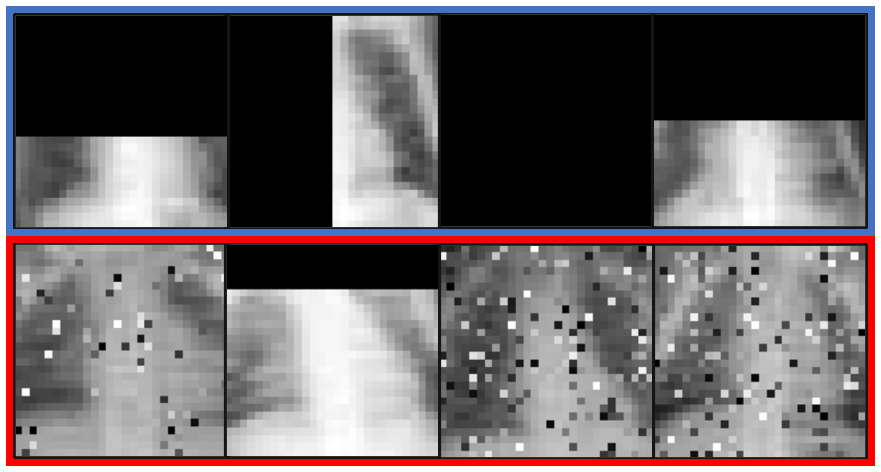}
    \caption{Controller prediction samples for the pneumonia detection task. \textbf{Blue}: samples predicted as having low task amenability; \textbf{Red}: samples predicted as having high task amenability. Low task amenability refers to controller predicted value below the 10-th percentile and high refers to above. More examples can be found with the open-source repository. (Colour figure)}
    \label{fig:my_label}
\end{figure}

\begin{figure}[!ht]
  \centering
  \subfloat[Prostate presence classification task]{\includegraphics[width=0.45\textwidth]{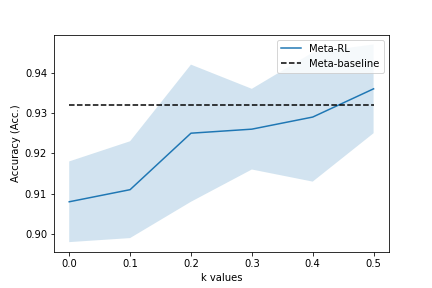}\label{fig:acc_ratio}}
  \hfill
  \subfloat[Prostate segmentation task]{\includegraphics[width=0.45\textwidth]{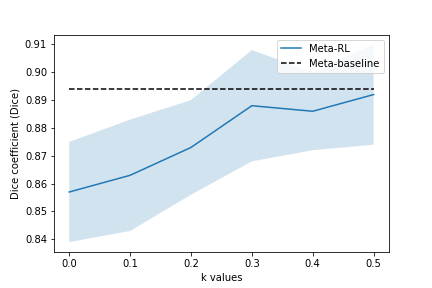}\label{fig:dice_ratio}}
  \caption{Plots of the task performance (in respective Acc. and Dice metrics) against the $k$ values with a rejection ratio set to 5\%. (Colour figure)}
  \label{fig:res_meta_IQA}
\end{figure}

\paragraph{Adaptability performance of meta-RL} The proposed meta-RL multi-environment framework was evaluated using the prostate classification and segmentation tasks and the results are summarised in Table \ref{tab:res_meta_IQA}. Additionally, a plot of performance against varying $k$ values is presented in Fig. \ref{fig:res_meta_IQA} for both tasks. Statistical significance was not found between the baseline and meta-RL for the prostate presence classification task, with $k$ values from $0.5$ to $0.2$ (\textit{0.10$<$p-values$<$0.23}), however, the baseline showed significantly higher performance compared to meta-RL for low k-values of $0.1$ and $0.0$ (\textit{p-values$<$0.01} for both). For the prostate segmentation task, no statistical significance was found between the baseline and meta-RL for $k$-values from $0.5$ to $0.3$ (\textit{0.07$<$p-values$<$0.17}) but the baseline performance was significantly higher than meta-RL for low $k$ values from $0.2$ to $0.0$ (\textit{p-values$<$0.01}). For the ablation study, the proposed meta-RL framework outperformed the meta-RL variant which had no environment-level separation, with statistical significance, for $k$ values from 0.0 to 0.4 (\textit{p-values$<$0.01}), for the classification task and for all $k$ values, for the segmentation task (\textit{p-values$<$0.03}). A significant difference was not observed for a high $k$ value of $0.5$ in the classification task (\textit{p-value}=0.06). In another ablation study, compared to a Reptile-omitted meta-RL variant, which achieved Acc.=$0.901\pm0.013$ and Dice=$0.851\pm0.013$ for the classification and segmentation tasks, respectively, the meta-RL framework showed higher performance, with statistical significance (\textit{p-values$<$0.01}), for both tasks for a $k$-value of $0.0$. For all other $k$-values no significant differences were seen between meta-RL and the its Reptile-omitted variant.

\begin{table}[!ht]
\centering
\caption{Comparison of holdout set results with a rejection ratio set to $0.05$ (Meta-RL)}
\begin{tabular}{|p{2.0cm}|p{0.6cm}|p{2.1cm}|p{2.1cm}|}
\cline{1-4}
\multicolumn{2}{|c|}{Tasks} & Prostate Classification (Acc.) & Prostate Segmentation (Dice)\\
\cline{1-4}
IQA Methods & k & Mean $\pm$ St.D. & Mean $\pm$ St.D.\\
\cline{1-4}
Meta-baseline & N/A & 0.932 $\pm$ 0.011 & 0.894 $\pm$ 0.016\\
\cline{1-4}
\multirow{6}{1.6cm}{Meta-RL} & 0.5 & 0.936 $\pm$ 0.012 & 0.892 $\pm$ 0.018\\
\cline{2-4}
& 0.4 & 0.929 $\pm$ 0.016 & 0.886 $\pm$ 0.014 \\
\cline{2-4}
& 0.3 & 0.926 $\pm$ 0.010 & 0.888 $\pm$ 0.020\\
\cline{2-4}
& 0.2 & 0.925 $\pm$ 0.017 & 0.873 $\pm$ 0.017\\
\cline{2-4}
& 0.1 & 0.911 $\pm$ 0.012 & 0.863 $\pm$ 0.020\\
\cline{2-4}
& 0.0 & 0.908 $\pm$ 0.010 & 0.857 $\pm$ 0.018\\
\cline{1-4}
\multirow{6}{1.6cm}{Meta-RL Variant} & 0.5 & 0.931 $\pm$ 0.015 & 0.884 $\pm$ 0.016\\
\cline{2-4}
& 0.4 & 0.920 $\pm$ 0.010 & 0.882 $\pm$ 0.021\\
\cline{2-4}
& 0.3 & 0.919 $\pm$ 0.013 & 0.882 $\pm$ 0.015\\
\cline{2-4}
& 0.2 & 0.916 $\pm$ 0.014 & 0.860 $\pm$ 0.014\\
\cline{2-4}
& 0.1 & 0.905 $\pm$ 0.014 & 0.858 $\pm$ 0.021\\
\cline{2-4}
& 0.0 & 0.896 $\pm$ 0.016 & 0.849 $\pm$ 0.017\\
\hline
\end{tabular}
\label{tab:res_meta_IQA}
\end{table}

\begin{figure}[!ht]
  \centering
  \subfloat[Prostate classification task]{\includegraphics[width=0.48\textwidth]{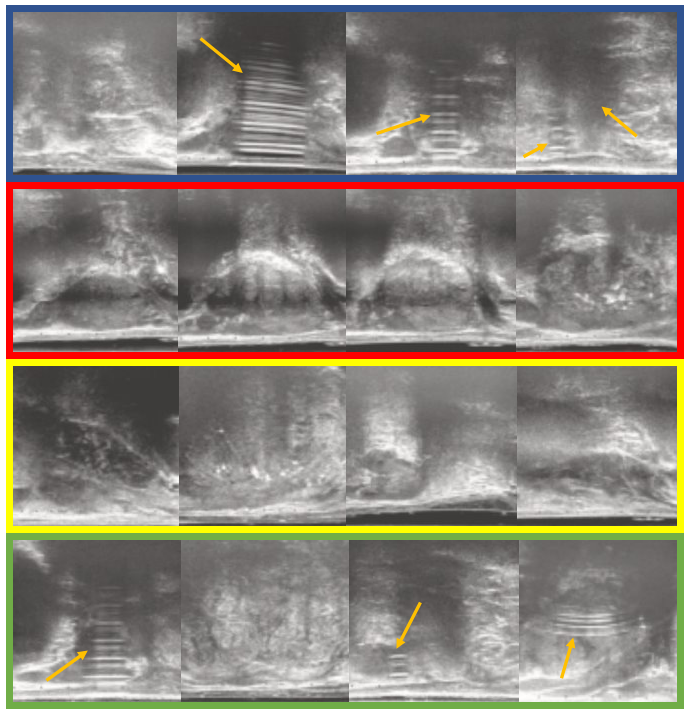}\label{fig:samples_class}}
  \hfill
  \subfloat[Prostate segmentation task]{\includegraphics[width=0.48\textwidth]{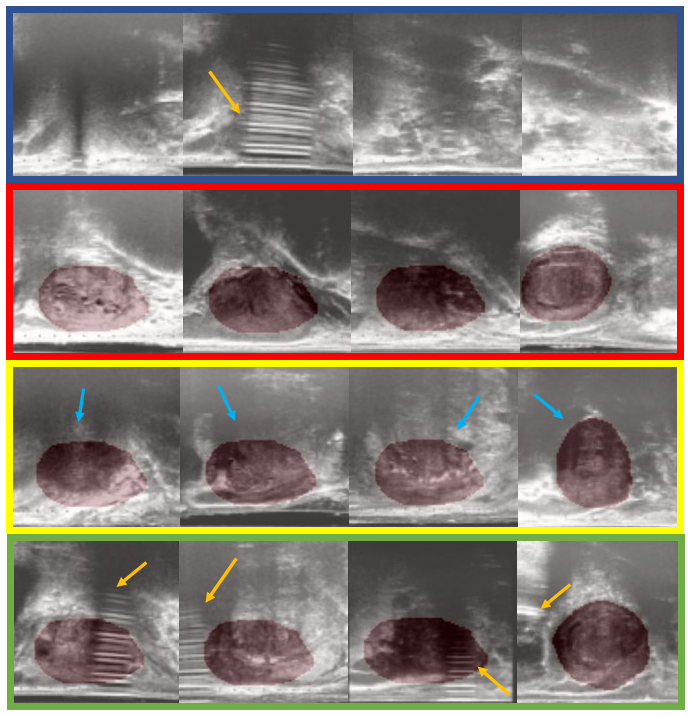}\label{fig:samples_seg}}
  \caption{Examples of controller selected and rejected images (rejection ratio=5\%) for both tasks for the multi-environment framework. \textbf{Blue:} rejected samples; \textbf{Red:} selected samples; \textbf{Yellow:} rejected samples despite no apparent artefacts or severe noise; \textbf{Green:} selected samples despite present artefacts or low contrast. \textbf{Orange arrows:} visible artefacts; \textbf{Cyan arrows:} regions where gland boundary delineation may be challenging. (Colour figure)}
  \label{fig:res:samples}
\end{figure}

\section{Discussion}

The results presented in Sect. \ref{sec:results} show that the proposed task amenability framework is able to offer an increased performance, compared to a non-selective baseline for all tested tasks which demonstrates the efficacy of the proposed approach. The three tasks include a range of clinical applications, including prostate presence classification, prostate gland segmentation and pneumonia detection, for surgical navigation, interventional guidance and diagnostic assistance. 

It was found particularly useful that task amenability can be learnt within the proposed IQA framework without any human labels of image quality, by using the weighted and selective reward formulations. While only the weighted reward formulation offers comparable performance to the fixed clean validation set reward formulation, which requires human labels of image quality, the selective formulation provides a means to specify a clinically desirable rejection rate. Furthermore, tuning this validation set rejection ratio parameter, $s_\text{rej}$,  for the selective formulation may be necessary in order to achieve performance comparable to the weighted- and fixed clean validation set reward formulations. In practice, we observed that increasing $s_\text{rej}$ beyond a certain point, the performance was significantly reduced which could potentially be due to over-fitting of the agent, caused by an increased exploration space as a result of the increased number of samples to be rejected from the validation set. This parameter could thus also impact the exploration-exploitation trade-off in training the proposed task amenability framework. Interestingly, the mean performance showed a small decrease after an initial rise, with increasing holdout set rejection ratio, for all tested reward formulations. The variance of predictions, possible over-fitting of the RL agent and the overall high quality of the datasets used (which could in turn limit the overall performance improvement), may offer some explanation for this observation, however, it still remains an open question.

The multi-environment framework, proposed to learn adaptable task amenability assessment, is potentially applicable to several scenarios, for example, learning an adaptable IQA definition over different target tasks or task predictor functions, besides different datasets. This multi-environment framework was investigated in the clinically relevant scenarios to equip adaptability, over different observer labels, to both the controller and task predictor functions. Adaptation to an expert labelling standard was achieved with 20-30$\%$ of the expert labelled data without any significant reduction in performance compared to using 100$\%$ of the expert labelled data for training. This meant that, for the classification and segmentation tasks, 1087 and 1634 expert-labelled images from 42 and 63 subjects (training and validation sets), respectively, were sufficient to achieve performance comparable to using 100$\%$ of the expert-labelled data for training. It is important to note that the multi-environment formulation also required non-expert-labelled data for training, however, these labels may be used to learn varying definitions of task amenability; further economic analysis of the use of non-expert data is beyond the scope of this work. The applicability of the framework, to equip adaptability to allow for efficient use of expert-labelled data, brings to light other potential use cases such as to equip adaptability across different target tasks to adapt IQA definitions to new tasks using limited labelled data which could be useful for several applications such as active learning, and particularly of interest in developing resource-constrained clinical applications.

\section{Conclusion}
This work introduces task amenability as an alternative to traditional subjective definition of image quality, especially for downstream machine learning tasks. We also propose a mechanism to efficiently adapt such RL-based IQA agents to new labelling standards. Learning a task-specific IQA or task amenability is useful for several applications such as for re-acquisition guidance for applications such as ultrasound where re-acquisition is inexpensive, for operator skill feedback and for meeting clinically defined accuracy requirements for downstream clinical tasks. The proposed multi- and single-environment frameworks promise a new method to learn such a task-specific IQA or task amenability. The proposed approach has shown its applicability with clinically relevant target tasks of prostate presence detection and gland segmentation, and pneumonia detection, with real clinical data from prostate cancer and pneumonia patients, respectively.

\section*{Acknowledgements}

This work is supported by the Wellcome/EPSRC Centre for Interventional and Surgical Sciences [203145Z/16/Z]; the International Alliance for Cancer Early Detection, an alliance between Cancer Research UK [C28070/A30912; C73666/A31378], Canary Center at Stanford University, the University of Cambridge, OHSU Knight Cancer Institute, University College London and the University of Manchester; EPSRC CDT in i4health [EP/S021930/1]; the Departments of Radiology and Urology, Stanford University; an MRC Clinical Research Training Fellowship [MR/S005897/1] (VS); the Natural Sciences and Engineering Research Council of Canada Postgraduate Scholarships-Doctoral Program (ZMCB); the University College London Overseas and Graduate Research Scholarships (ZMCB); GE Blue Sky Award (MR); and the generous philanthropic support of our patients (GAS). Previous support from the European Association of Cancer Research [2018 Travel Fellowship] (VS) and the Alan Turing Institute [EPSRC grant EP/N510129/1] (VS) is also acknowledged.

\bibliographystyle{model2-names.bst}\biboptions{authoryear}
\bibliography{bibliography}

\end{document}